# A Storage Advisor for Hybrid-Store Databases

Philipp Rösch
SAP Research, SAP AG
Chemnitzer Str. 48
01187 Dresden, Germany
philipp.roesch@sap.com

Lars Dannecker
SAP Research, SAP AG
Chemnitzer Str. 48
01187 Dresden, Germany
lars.dannecker@sap.com

Gregor Hackenbroich
SAP Research, SAP AG
Chemnitzer Str. 48
01187 Dresden, Germany
gregor.hackenbroich@sap.com

Franz Färber
SAP AG
Dietmar-Hopp-Allee 16
69190 Walldorf, Germany
franz.faerber@sap.com

## ABSTRACT

With the SAP HANA database, SAP offers a high-performance in-memory hybrid-store database. Hybrid-store databases—that is, databases supporting row- and column-oriented data management—are getting more and more prominent. While the columnar management offers high-performance capabilities for analyzing large quantities of data, the row-oriented store can handle transactional point queries as well as inserts and updates more efficiently. To effectively take advantage of both stores at the same time the novel question whether to store the given data row- or column-oriented arises. We tackle this problem with a storage advisor tool that supports database administrators at this decision. Our proposed storage advisor recommends the optimal store based on data and query characteristics; its core is a cost model to estimate and compare query execution times for the different stores. Besides a per-table decision, our tool also considers to horizontally and vertically partition the data and manage the partitions on different stores. We evaluated the storage advisor for the use in the SAP HANA database; we show the recommendation quality as well as the benefit of having the data in the optimal store with respect to increased query performance.

## 1. INTRODUCTION

Recent studies reveal a rapid growth of data volumes in data warehouse systems [16]. At the same time, there is an increasing need for interactively analyzing these data to utilize hidden information. These two factors gave rise to a revival of column-oriented databases (column stores). Those column stores are optimized for data-intensive analysis queries (OLAP - Online Analytical Processing). On the other hand, there are traditional data stores that are oriented row-wise instead. Row-store technology is widely used and highly advanced. Especially, row stores are (much) better suited in transactional scenarios (OLTP - Online Transactional Processing) with many updates and inserts of data as well as point queries.

A current trend in databases is to leverage the advantages of both techniques by comprising both a row and a column store (e.g., see [13]). Such hybrid-store databases offer high-performance transactions and analyses at the same time. They provide the capability to increase the efficiency of operational BI for enterprises, where in most cases mixed workloads comprising OLTP and OLAP queries are used.

To effectively take advantage of the different storage formats of hybrid-store databases one carefully has to decide whether to store the data row- or column-oriented. This poses a novel question for database optimization: Which data should be managed in which store? This decision is not trivial and support for the database administrator would be of great value. However, hybrid-store databases are rather new and most of their tools, optimizations, and database administrator support utilities are focusing on individual parts instead of considering them jointly. Hence, currently there is insufficient support for this new dimension of database optimization.

We address this point with a storage advisor tool for hybrid-store databases. Besides recommending the store for complete tables the storage advisor also considers to partition the tables both horizontally and vertically among the two stores to allow more fine-grained decisions, and thus, to further optimize performance. As a result, the proposed storage advisor relieves the database administrator of the decision where to store the data in a hybrid-store database. On the one hand, this saves time and money as the administrator now can focus on other ways to improve the performance. On the other hand, the advisor may result in better decisions compared to the experiences or speculations of the database administrator, which in turn yields better performance and faster results.

Our work focuses on implementing the storage advisor for the SAP HANA database [8], which is an in-memory hybrid-store database. In-memory databases are a new development where all the data is stored in the computer's main memory. As a consequence, no hard disc access is required at runtime allowing for significantly faster query processing. However, although developed and evaluated for the SAP HANA database, the main principle of the storage advisor proposed in this paper can also be adapted for use in other hybrid-store database systems.

Summarizing, in this paper we make the following contributions:

- In Section 3, we introduce a cost model as a basis for the row/column store decision. This cost model allows for decisions on table level but also on partition level; for that, it takes joins between tables of different stores into account.

- In Section 4, we present some implementation details of the storage advisor tool. We show two different working modes (offline and online) and describe the recommendation process.

- With a detailed evaluation in Section 5, we demonstrate the advantages of the storage advisor. We analyze the accuracy of the cost estimation and the runtime benefits of mixed workloads when using the storage advisor.

In Section 6, we discuss related work, and we conclude the paper with a summary and future work in Section 7.

## 2. ROW, COLUMN, AND HYBRID STORES

In this section, we briefly summarize the specifics of the different stores of a hybrid-store database with respect to workload and data characteristics. Additionally, we provide some details about the SAP HANA database, the hybrid-store database used in this paper.

*Row Store Specifics.* In a row store, data is managed on tuple level, i.e., all the different attribute values of the individual tuples are stored subsequently. This way, accessing (a few) complete tuples can be done very efficiently. Clearly, this is beneficial for retrieving tuples but also for updates. Additionally, inserts can be efficiently realized by appending the new tuples to the row store.

*Column Store Specifics.* Contrary, in a column store the data is managed on column or attribute level, i.e., all the values of the individual attributes are stored subsequently. That way, accessing all values of a column—as typically done for aggregates—can be done very efficiently. Moreover, the higher similarity of values within a column compared to the similarity across columns allows speeding up query processing by using compression. This, in turn, entails some extra cost for updating and inserting tuples. Also, accessing multiple attributes requires to join the requested attributes.

Consequently, different query types benefit from different alignments of the data in the memory, i.e., from different stores, as illustrated in Figure 1. Hence, as a consequence of the store's specifics, some workload characteristics turn out to be decisive for the appropriateness of either of the two stores. For example, the number of requested attributes, the characteristics of the aggregates, and the rate of inserts and updates of a workload have to be considered when selecting the store. Also, data characteristics such as the number of distinct values as well as the frequency skew become relevant in the context of compression. We address these decision criteria in more detail in Section 3.

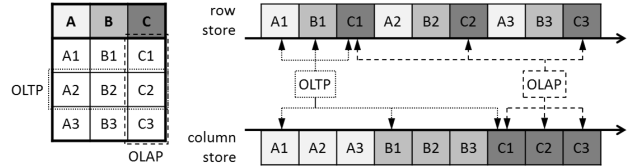

**Figure 1: Memory alignment and resulting access patterns for row and column stores**

*SAP HANA Database*

As mentioned in the introduction, our proposed storage advisor is implemented for the SAP HANA database, SAP's in-memory hybrid-store database. The SAP HANA database is a full database management system with a standard SQL interface, transactional isolation and recovery, and high availability. To natively take advantage of massively parallel multi-core processors, SAP HANA manages the SQL processing instructions into an optimized model that allows parallel execution and scales extremely well with the number of cores. Additionally, by managing all the data in main memory, materialized views become superfluous; aggregates are computed on the fly and the expensive selection and maintenance of materialized views are omitted.

With the hybrid-store architecture, the SAP HANA database is optimized for both transactional and analytical scenarios. Its row store is highly optimized for inserts, updates, and row-by-row processing while the column store is extremely fast at computing aggregates by using compression (e.g., [12]). The two stores of the hybrid-store database are transparent for the user when querying the data.

## 3. STORAGE ADVISOR COST MODEL

With the storage advisor for hybrid-store databases, we provide a tool that automatically decides whether to store given data row- or column-oriented. In the most basic way, the storage advisor provides its recommendations on table level (see Section 3.1). That is, for each table in the database the storage advisor recommends the appropriate store. However, as usually different parts of a table have different access patterns a more fine-grained decision may allow for even larger performance gains. Concretely, tables may be split into different parts managed in different stores. Hence, our tool also addresses partitioning of tables to further increase the efficiency of hybrid-store databases, as discussed in Section 3.2.

### 3.1 Recommendation on Table Level

To recommend the optimal store the storage advisor has to estimate the costs that would be incurred by storing the data in the respective store. In our case, the cost is given by the query runtime, i.e., the goal of the storage advisor is to assign the given data to the row store and to the column store in such a way that the runtime of a given workload is minimal. As a constraint, to ensure consistency and minimize memory consumption, data may only be stored in one or the other store but no in both.

For the estimation and comparison of query runtimes the storage advisor utilizes a cost model that takes both query and data characteristics into account. The actual contributing query and data characteristics depend on the query type as shown below. Candidates for query characteristics are the number of aggregates and the applied aggregation function for OLAP queries or the number of affected rows for OLTP queries. Candidates for data characteristics are the data volume and the data types.

Detailed preceding analyses revealed that the characteristics influence the query runtimes independently from each other. As a result, our cost model is designed as follows:

$$Costs = BaseCosts \cdot QueryAdjustment \cdot DataAdjustment$$

Basically, this means that we have some base costs for a default setting and adapt them to the actual query and data characteristics. The query type of the current query determines the base costs and identifies query and data adjustments to be applied. That is, starting with base costs that are specific to the given query type, we apply query and data adjustments relevant for that query type. In more detail, both the query and the data adjustments are not atomic but are made up of different components according to the data and query characteristics mentioned above. We discuss the adjustments for different query types in the following subsections.

### Aggregation Queries

For aggregation queries, typical characteristics for the query adjustment are the aggregation function and the grouping; characteristics for the data adjustment are the data type, the data volume and the compression rate of the data. As an example, for a simple query that computes the sum of an attribute the cost model evolves to

$$\begin{aligned} Costs &= BaseSUMCosts \cdot c_{groupBy} \cdot \\ & c_{dataType} \cdot f_{\#rows} \cdot f_{compression} \,. \end{aligned}$$

That is, we start with query-specific base costs ($BaseSumCosts$) and adapt it to the grouping ($c_{groupBy}$), to the data type of the aggregated attribute ($c_{dataType}$), to the number of rows to be read ($f_{\#rows}$), and to the compression rate of the data ($f_{compression}$). The adaptation to the grouping is a multiplication with a constant value. The reason is that in our system only the existence of the grouping is important and not the size or the number of groups. Similarly, also the adaptation to the data type is realized by a multiplication with constant value. The remaining adaptations are functions of the number of tuples and the compression rate, respectively.

So far, we have a cost model that considers the query type as well as query and data adjustments. Still, as the row and the column store behave differently on changes to data and query characteristics all the adjustment functions and constants are specific to the store. Consequently, we use store-specific base costs and adaptations, like $BaseSUMCosts^{RS}$ and $c_{grouping}^{RS}$ for the row store, and $BaseSUMCosts^{CS}$ and $c_{grouping}^{CS}$ for the column store.

Returning to our example, let the query compute the overall sum, i.e., no group by. Further, let the data type of the aggregated attribute be `Double`, the size of the data be 1000 tuples, and the compression rate be 0.7. If we now want to estimate the runtime for executing the query on the row store and include the parameters and parameter values into the cost model, we get

$$\begin{aligned} Costs &= BaseSUMCosts^{RS} \cdot c_{NoGroupBy}^{RS} \cdot \\ & c_{Double}^{RS} \cdot f_{\#rows}^{RS}(1000) \cdot f_{compression}^{RS}(0.7) \,. \end{aligned}$$

The required data characteristics are efficiently retrieved from the system catalog; for the storage advisor, we extended the system catalog to maintain the relevant statistics, like here, the compression information.

An extension of the query by a second aggregate—here, the average of an integer column—would result in an extension of the cost model to

$$\begin{aligned} Costs &= (BaseSUMCosts^{RS} \cdot c_{Double}^{RS} + \\ & BaseAVGMCosts^{RS} \cdot c_{Integer}^{RS}) \cdot \\ & c_{NoGroupBy}^{RS} \cdot f_{\#rows}^{RS}(1000) \cdot f_{compression}^{RS}(0.7) \,. \end{aligned}$$

That is, the additional aggregate adds another base cost term including its adjustment to the data type to the cost model prior to the adjustments to the remaining query and data specifics.

### Point and Range Queries

For point and range queries, the costs are estimated as follows:

$$\begin{aligned} Costs &= BaseSelectCosts \cdot f_{\#selectedColumns} \cdot \\ & f_{selectivity} \,. \end{aligned}$$

That is, starting from the base costs for selection queries, we adjust the cost estimate by the number of selected columns and the selectivity. The former is only relevant for data residing in the column store as here, the tuple reconstruction results in increasing costs; for the row store, $f_{\#selectedColumns}$ is just a constant function. Additionally, $f_{selectivity}$ is a function of both the selectivity of the query and the availability of an index. For the row store, $f_{selectivity}$ is a linear increasing function if an index is available while it is constant if no index may be used—in this case, a table scan is executed. On the other hand, for the column store $f_{selectivity}$ is a linear increasing function in any case as the data dictionary used for compression provides an implicit index. Hence, as above, the adaptation functions are specific to the stores the data reside in.

### Inserts and Updates

For insert and update queries, the costs are estimated by

$$\begin{aligned} Costs &= BaseInsertCosts \cdot f_{\#rows} \\ Costs &= BaseUpdateCosts \cdot f_{\#affectedColumns} \cdot \\ & f_{\#affectedRows} \,. \end{aligned}$$

For insert queries, we start with base costs and adjust it to the number of rows in the table. This adjustment is required, as uniqueness of the primary key and other unique attributes has to be verified before inserting the new tuple. For update queries, the cost estimation is rather similar to that of range queries: With $f_{\#affectedColumns}$, we take account of tuple reconstruction efforts for column store data, and $f_{\#affectedRows}$ basically reflects the selectivity of the query.

*Join Queries*
So far, our discussion focused on single, isolated tables. However, as many queries involve multiple tables it is important to take joins between tables into account. While it might be beneficial to have one table in the row store and another one in the column store when queried independently, it may be better to move both tables to the same store when they are often used for joins. This saves the conversion of the different memory layouts and allows for faster joins. Fortunately, this issue is automatically addressed by estimating the query response times for join queries for the different store combinations. For the join of two tables this means four estimates compared to two estimates for the single table case; yet, as estimation can be done very efficiently this is a negligible overhead.

The estimation for join queries requires to adapt and extend the cost model. Now, the base costs are determined by multiple tables and are adjusted to the characteristics of all tables. Extending our first example query above by a join with a table of 100000 rows and estimating the costs for managing the first table in the row store (RS) and the second one in the column store (CS), we get

$$\begin{aligned} Costs \quad = \quad & BaseSUMCosts^{RS,CS} \cdot c_{NoGroupBy}^{RS} \cdot c_{Double}^{RS} \cdot \\ & f_{\#rows}^{RS}(1000) \cdot f_{\#rows}^{CS}(100000) \cdot \\ & f_{compression}^{RS}(0.7) \cdot f_{compression}^{CS}(0.5) \,. \end{aligned}$$

Here, we see the base cost is determined by both stores as well as adjustments to data characteristics for both tables.

For all the cost estimation functions given above, the base costs and the adjustment functions $f$ have been determined by detailed performance analyses of the different database operations. As stated above, the adjustment functions are independent from each other. Moreover, most of these functions are simple linear functions (e.g., $f_{\#rows}$), piecewise linear functions (e.g., $f_{compression}$) or even constants (e.g., $c_{dataType}$). Both, the independence and the simplicity of the adjustment functions make the estimation of query runtimes highly efficient.

Now, having determined the cost model, the data characteristics in the system catalog, and the query characteristic from the current workload the storage advisor can estimate and compare the workload runtimes for managing the tables in the row store and in the column store and provide a recommendation for the optimal store for the individual tables. Details of this process are given in Section 4.

### 3.2 Store-aware Partitioning

With the approach presented above, the storage advisor provides its recommendation on table level, that is, for each table we get one preferred store for all its tuples and attributes. However, as usually different parts of an individual table have different access patterns a more fine-grained decision allows for further performance gains. To leverage this fact we consider splitting the tables into different parts and manage them in different stores. In other words, we support to horizontally and vertically partition the tables and put the partitions into the more beneficial store, respectively.

*Horizontal Partitioning.* For the horizontal partitioning, a table is split into disjoint sets of rows. This partitioning scheme potentially increases the efficiency of query processing when one subset of the tuples is (mainly) used for analytical queries while the other subset is subject to frequent updates or point queries. In this setting, historic tuples could be stored in the column-store partition for fast analyses while current and newly arriving tuples are stored in the row-store partition, which allows for faster inserts. In certain intervals, data is moved from the row-store partition to the column-store partition. For queries addressing all the data of the table, a union of both partitions is executed. Another scenario for horizontal partitioning is to store tuples that are no longer used for analysis in a row-store to increase the efficiency of analytical queries on the column-store partition. The horizontal partitioning scheme is sketched in Figure 2.

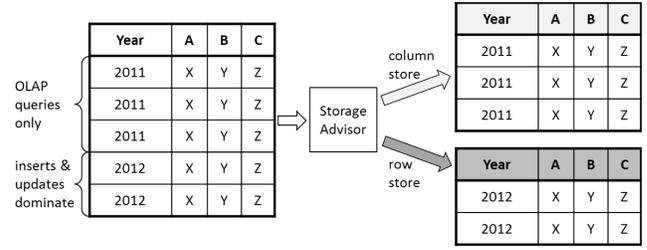

Figure 2: Horizontal partitioning

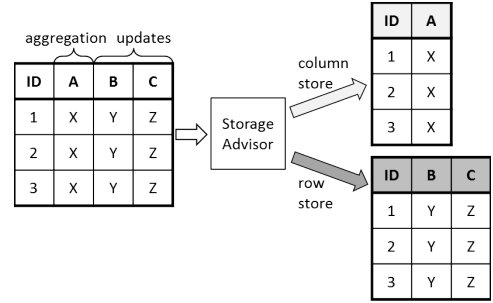

Figure 3: Vertical partitioning

*Vertical Partitioning.* For the vertical partitioning, a table is split into sets of columns (or attributes). In contrast to the horizontal partitioning, the partitions are not disjoint but all contain the primary key attributes. This technique is beneficial when a subset of the attributes is mainly subject to analytical queries (like prices or quantities) while other attributes are often modified (like status on shipment or payment). In such a scenario, the partition with the typically aggregated attributes is stored in the column store while the other attributes are managed in the row store. For queries addressing all the data of the table, the partitions have to be joined. Figure 3 illustrates the vertical partitioning scheme.

To support highly efficient inserts, updates of individual attributes, and high-performance analyses at the same time, the partitioning of the tables is not restricted to either horizontal or vertical but both partitioning schemes may be applied at the same time. As an example, the historic tuples of a table can be split vertically as shown in Figure 3 as still updates on status attributes may occur while newly arriv-

1751

ing tuples are stored as a whole in an individual horizontal partition as shown in Figure 2.

*Recommend Partitions.* The recommendation of partitions is a significantly harder problem as each query in the workload may have a different optimal partitioning. Even determining the optimal partitions for a single query is prohibitive expensive as the number of possible partitions is extremely high. For this reason, we propose a simplified, heuristic approach. For each table, we consider (up to) two horizontal and (up to) two vertical partitions. More concretely, we first consider splitting the table horizontally to allow for high-performance analysis of historic data in the column store and fast inserts of new data into the row store. For the partition with the historic data, we additionally consider a vertical split into (rather) static attributes for the column store and often modified attributes for the row store. The decision whether or not to split a table and which tuples and attributes to put into which store, i.e., where to split the table, is based on an analysis of the workload and works as follows:

**Horizontal partitioning**

- **Get fraction of insert queries** to determine if a partition for inserts is meaningful. This fraction can easily be derived from the queries in the workload; if it is sufficiently high a row-store partition for newly arriving tuples will be recommended.

- **Get tuples that are frequently updated** as a whole, that is, updates that are addressing many attributes or whose predicates include many attributes. To determine those tuples we could either use extended statistics (for details see Section 4) or estimate those tuples based on the queries and standard table statistics. In case there are such tuples the storage advisor will recommend to put them into a row-store partition.

**Vertical partitioning**

- **Get OLTP attributes**, i.e., attributes that are mainly and often used for updates or point queries rather than analyses. Those attributes can easily be determined from the queries in the workload. If there are such attributes, the storage advisor recommends to create a respective row-store partition.

Based on this heuristic approach, the storage advisor will recommend partitions for the respective tables. In the next section, we will present some details on the implementation of the storage advisor.

## 4. IMPLEMENTATION

We implemented the storage advisor in the SAP HANA database as a tool to support the database administrator. Basically, the storage advisor supports two working modes—the offline and the online mode.

*Offline Mode.* The offline mode is mainly used to provide initial recommendations. In this mode, the input for the storage advisor are the given database schema, basic table statistics for the data characteristics, and recorded or expected workload information, see Figure 4. As a result, the storage advisor recommends an (initial) storage layout (and a system-specific cost model as discussed below). An advantage of the offline mode is that the input information is relatively cheap and simple to acquire. However, the recommended storage layout is based on some approximations and, thus, may not be optimal. Additionally, as soon as the workload evolves and deviates from the recorded or expected workload the storage layout may get more and more inappropriate.

*Online Mode.* The online mode is used to check the storage layout at runtime of the database system. Here, the input for the storage advisor are the given database schema, extended table statistics, and detailed workload statistics, see Figure 4. Contrary to the offline mode, the storage advisor now can continuously recommend beneficial storage layout adaptations. Also, as the recommendations are based on more detailed information the results will be closer to the optimal storage layout. The disadvantage of the online mode, however, is that recording the required statistics is an expensive task. Examples for extended workload statistics are information about the number of inserts per table, the number of updates and aggregates per attribute or the number of joins between tables.

Clearly, the two modes are not mutual exclusive but are typically combined. Starting with an offline recommendation for the initial storage layout, we can record extended table and workload statistics to enable online storage layout adaptations. Figure 4 illustrates that in both the offline and the online mode the storage advisor consumes data characteristics and schema information. The difference is in the used workload information. For the offline mode, a recorded or expected workload is used. In contrast, in the online mode live statistics about the executed queries are recorded. As a result, in the online mode the storage advisor is typically more accurate as it covers the actual and most current query executions. It is important to note that in the online mode the storage advisor uses the cost model that is initialized in the offline mode as discussed below.

*Recommendation Process.* The process of the storage advisor to recommend a storage layout is shown in Figure 5. As the first step, the storage advisor initializes the cost model, that is, based on some representative tests the base costs and the adjustment functions (see Section 3) are set to reflect the current system's hardware settings and system configurations. Next, the cost model is used to recommend an initial storage layout in the offline mode. After that, the storage advisor runs in the online mode. That is, extended workload and table statistics are recorded and, in certain time intervals, the storage advisor re-evaluates the storage layout based on the current workload statistics and recommends adaptations if required. Optionally, to also keep track of changes in hardware or system settings the storage advisor may re-initialize the cost model from time to time. If there are (noticeable) changes, the storage advisor re-evaluates the storage layout and recommends the respective adaptations.

*Application of the Recommendations.* The recommendations of the storage advisor are typically presented to the database administrator. This also includes the respective statements to move the data into the recommended store. Hence, the database administrator can easily ask the stor-age advisor to apply the recommended storage layout. Alternatively, the recommended storage layout could also be applied automatically. This completely eliminates the interaction with the database administrator. However, as the adaptation of a storage layout impacts the system's performance and availability of the respective tables this option should be applied with care.

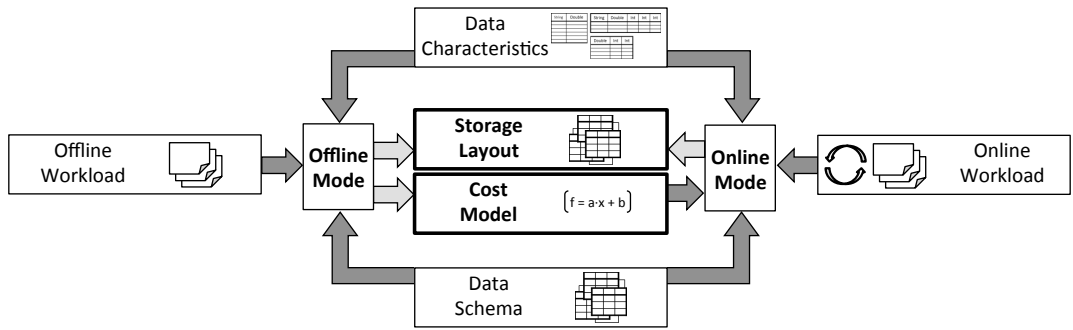

Figure 4: Inputs and outputs for the storage advisor working modes

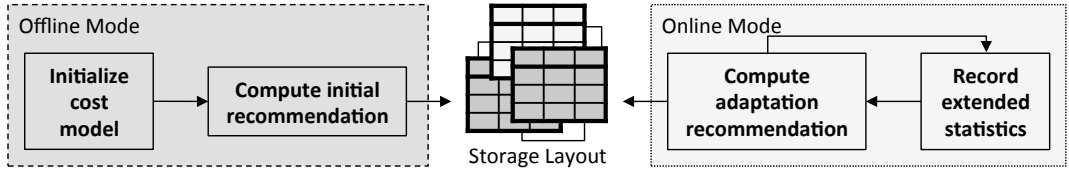

Figure 5: Store recommendation process

*Store-aware Partitioning.* As already described, the storage advisor also considers the partitioning of tables among both stores (compare Section 3.2). This, however, entails that the query processing has to be aware of the applied partitioning. As users are typically not aware of the partitioning the query rewriting must be realized automatically and transparently to the user. Therefore, we extended the system catalog; now, for each table, there is an annotation that describes the partitioning of the storage advisor. For incoming queries, this annotation is evaluated and if required, the query is rewritten accordingly.

## 5. EVALUATION

In this section, we evaluate the proposed storage advisor. We conducted several experiments to analyze the accuracy of the cost model and the benefits of the recommended storage layout in terms of reduced workload runtimes. We evaluated the decision on table level and on partition level and compared the different approaches.

*Summary of Results*

The most important results of our evaluation are as follows:

- With the proposed cost model, we can estimate query runtimes highly accurate. This is an important basis for recommending the optimal store.

- For the recommendation on table level, the storage advisor is able to identify the optimal store depending on the workload. As the workload mix strongly influences the optimal store this may significantly reduce the workload runtime.

- Additionally, the optimal store-aware partitioning can further reduce the workload runtimes. We show this both for horizontal and vertical partitioning.

- Finally, with a combination and comparison of the approaches we show the benefit of the optimal storage layout for a TPC-H like scenario. Here, table-level and partitioning recommendations can reduce workload runtimes by 40% and 65%, respectively compared to managing all the tables in one of the two stores.

### 5.1 Settings

Our performance measurements were conducted on a machine with two 8-core CPUs operating at 2.53 GHz and 64 GB RAM running SLES11. The given runtimes are averaged over several runs.

For most of the experiments, we carefully generated different data sets and workloads to analyze the impact of different data and query characteristics. For our final experiment, we used the well-known TPC-H (www.tpc.org/tpch) data and ran a mixed workload to also evaluate the storage advisor in a common enterprise setting.

### 5.2 Estimation Accuracy

In the first series of experiments, we evaluated the accuracy of the cost estimation of the storage advisor. Here, we varied both data and query parameters and compared the estimates of the storage advisor with the actual runtimes. Below, we show the results for the scale of the data and the scale of aggregates in a query.

*Data Scale.* To analyze the accuracy for different data scales we ran a constant aggregation query against a table for different data volumes ranging from $2m$ to $20m$ tuples;

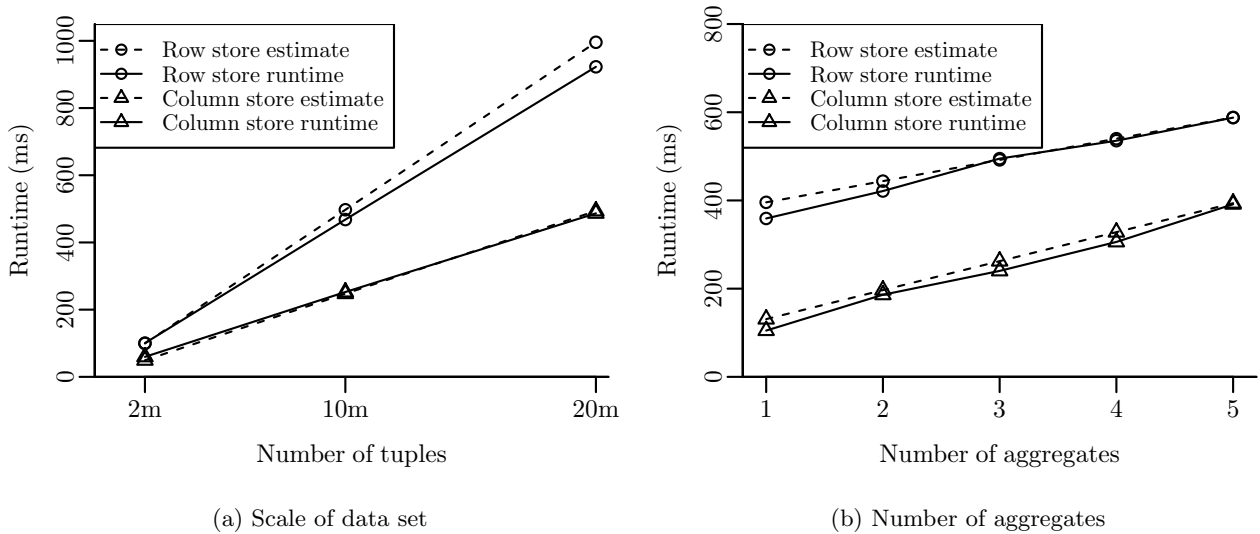

(a) Scale of data set  (b) Number of aggregates

**Figure 6: Accuracy of the runtime estimation**

the table consisted of 30 attributes (ID and several keyfigures, filter attributes, and group-by attributes). The results are shown in Figure 6(a). As can be seen, both stores show a linear trend. Besides the actual runtimes, the plot shows the estimates from the storage advisor. Especially for the column store, the estimates (dashed lines) are very close to actual runtimes (continuous lines).

*Query Scale.* In the next experiment, we varied the number of aggregates of the aggregation query. We used the same table setting as above; the table had a data volume of $10m$ tuples. Figure 6(b) shows the results. Again, we can see a linear trend for both stores. Also, the estimates are again very close to the actual runtimes.

*Summary.* With these experiments, we briefly showed the high accuracy of the cost estimation for different settings. Clearly, the accuracy of the cost estimation is the basis for meaningful recommendations of the storage advisor. Recall, the cost model uses adjustment functions $f$ to compute the costs for specific data and query settings (see Section 3). The evaluation of the accuracy presented above also shows the shape of the respective adjustment functions and also exemplifies the linearity of some of the adjustment functions.

## 5.3 Store Recommendation

Our next experiments focus on the accuracy of the recommendations of the storage advisor. We compare the runtimes of the store recommended by the storage advisor with the runtimes achieved when using only the row or column store irrespective of the data and query characteristics.

*Recommendation on Table Level*

In the basic case, the storage advisor recommends the store on table level, i.e., for each table and all its tuples and attributes the storage advisor recommends the optimal store. In the following, we evaluate the effectiveness of the recommendations.

*Single Table.* We first evaluate the recommendation accuracy for a workload querying a single table only. We used the same table as in the experiments above with a data volume of $10m$ tuples. For this table, we generated a workload consisting of different OLAP and OLTP queries. For the OLAP queries, we generated queries using different aggregation functions on different keyfigures; for OLTP queries, we generated a mix of insert and update queries. Next, we used the OLAP and OLTP queries to assemble different workloads. We varied the ratio of OLAP and OLTP queries in the workload and asked the storage advisor for the preferred store. Figure 7(a) shows the runtimes for different workloads consisting of 500 queries on the two stores as well as the runtime achieved by using the recommended store. As can be seen, both stores show a linear increase of workload runtimes for increasing OLAP fractions though the row store starts with shorter runtimes but shows a stronger increase. Hence, for workloads with very few OLAP queries (0% − 2,5%) the row store is the better option while for higher OLAP fractions the column store becomes the preferred store. In the figure, the continuous line shows the runtime for the workload when using the store recommended by the storage advisor, which is very close to the optimal case. Only when the runtimes for both stores are very similar (in the figure, for the 2.5% OLAP fraction) the storage advisor does not recommend the optimal store. However, in this case the overhead for not using the optimal store is very low.

*Joins.* Next, we evaluated the recommendation accuracy for workloads including join queries. For the data setting, we used a typical star schema. Based on preceding measurements, we always put the (small) dimension table into the row store and used the storage advisor to determine the optimal store for the fact table. As for the single table case, we used a fix data setting and generated workloads with different ratios of OLAP and OLTP queries. The fact table consisted of 10 attributes and $20m$ tuples while the dimension table contained 1000 tuples with 6 attributes. The

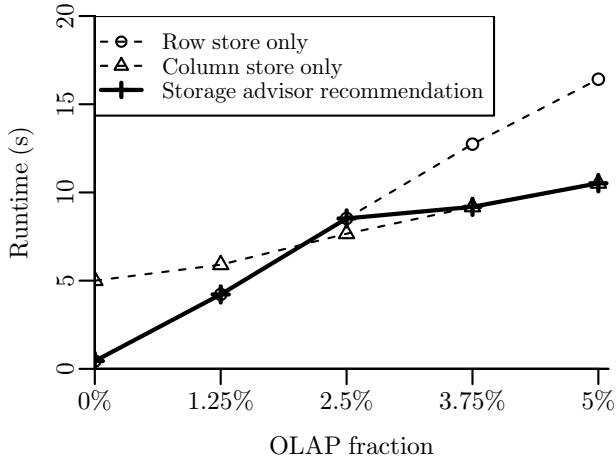
(a) Single-table queries

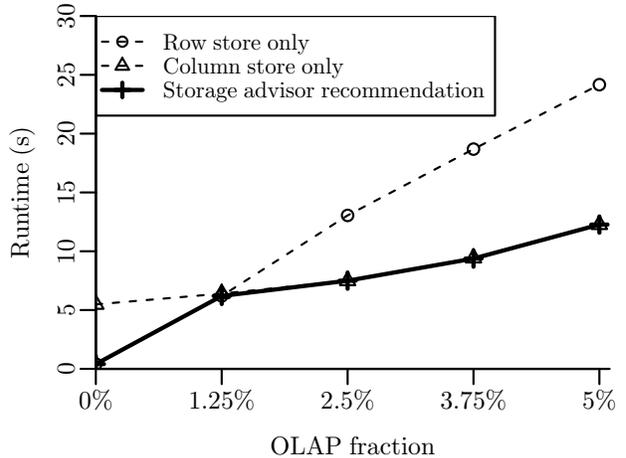
(b) Join queries

Figure 7: Recommendation quality

OLAP queries aggregated different keyfigures using different aggregation functions and grouped the data based on different attributes from the dimension table; the OLTP part of the workload updated tuples of the fact table and inserted new tuples into the fact table. As shown in Figure 7(b), the results are very similar to the single table case. However, now the storage advisor actually recommends the optimal store, and also, the optimal OLAP fraction for switching the store is lower.

*Summary.* The evaluation results presented above clearly show the advantages of choosing the right store already on table level—the query mix in the workload heavily influences the optimal store for the data, that is, managing the data in the wrong store results in significantly larger runtimes. Even more important, our proposed storage advisor precisely estimates the runtimes for both stores and thus, recommends (almost) always the optimal store resulting in (close to) minimal runtimes for all the different workloads.

### Recommendation on Partition Level

The next series of experiments evaluates the impact of the store-aware partitioning on the runtime of a workload.

*Horizontal Partitioning.* In the first experiment, we analyzed the benefit of horizontal partitioning for a fixed workload. To this end, we used a table as for the previous experiments and generated a mixed workload of 500 queries with an OLAP fraction of 5% and update queries addressing 10% of the data (referred to as OLTP data). The storage advisor recommended to put these 10% of the data in a row-store partition while managing the remaining data in the column store (see heuristic given in Section 4). Now, we varied the partitions by putting different amounts of data into the row-store partition (thus, ignoring the recommended partitioning) and ran the workload on that partitioning setup. That way, we analyzed the impact of the partitioning on the workload runtime as well as the appropriateness of the recommended partitioning. We started with putting all the data into the column store and subsequently moved more and more of the OLTP data to the row store; for row-store partitions containing more than 10% of the data we moved some additional random non-OLTP data to the row store. As can be seen from Figure 2, having exactly the recommended 10% of OLTP data in the row store results in the lowest runtime of the workload; increasing or decreasing the size of the row-store partition results in a (rather) linear increase of the workload runtimes.

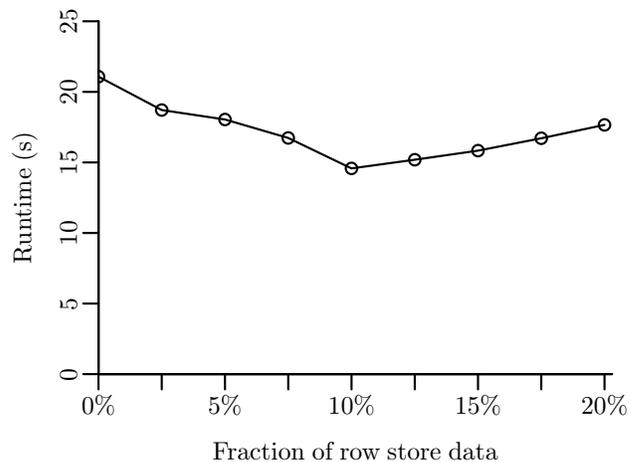

Figure 8: Runtime of workload for different horizontal partitionings

*Vertical Partitioning.* Next, we evaluated the vertical partitioning both for an OLAP and an OLTP setting. In the OLAP setting, the table consisted of 10 keyfigures, 8 group-by attribute, and only 2 attributes used for selections or updates; in the OLTP setting, the table had 18 attributes

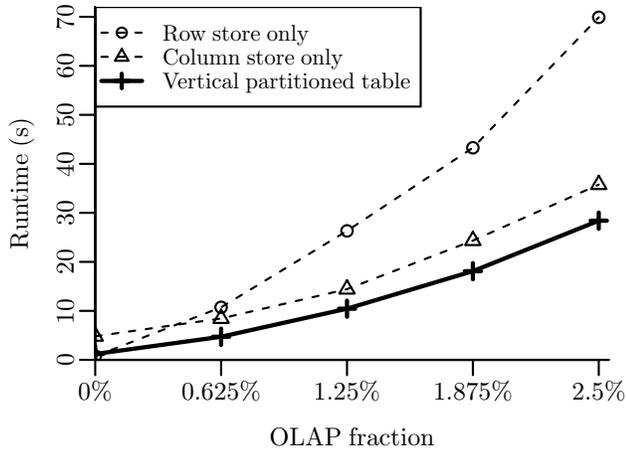

(a) OLAP setting

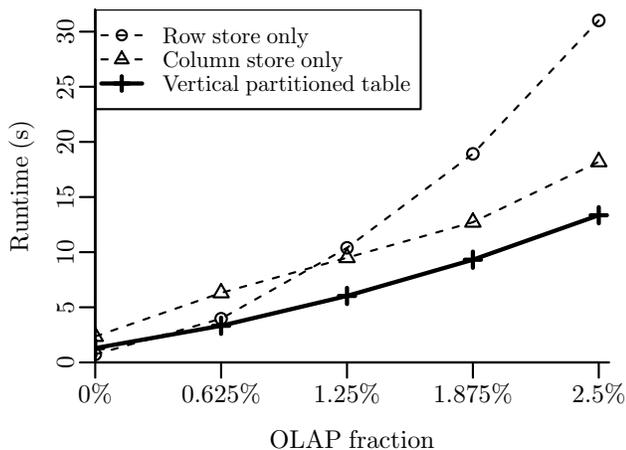

(b) OLTP setting

**Figure 9: Benefit of vertical partitioning on workload runtime**

used for selections and updates, and only 1 keyfigure and 1 group-by attribute. Accordingly, the workloads for both settings differ to fit to the table structure. As a result, after analyzing the workload the storage advisor recommended to put the respective OLAP attributes (keyfigures and group-by attributes) into the column store and the remaining attributes into the row store. Now, as for the evaluation of the decision on table level we generated different workload mixes by varying the fraction of OLAP queries in the workload. For each setting, we ran the workloads on a row-store table, on a column-store table and on a vertically partitioned table as recommended by the storage advisor. Figure 9(a) shows the runtimes for the OLAP setting and Figure 9(b) for the OLTP setting. In both settings, the vertical partitioning allows for lower runtimes compared to the unpartitioned cases except for the pure OLTP workloads (i.e., 0%

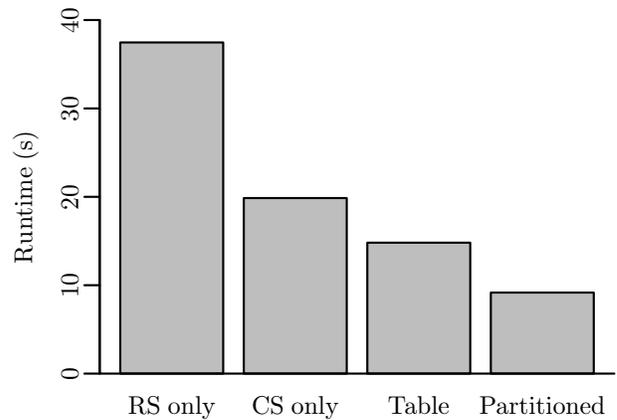

**Figure 10: Comparison of decisions on different levels**

OLAP fraction). Generally, it behaves similar to the column store settings but with a certain runtime decrease. Note that we focused only on vertical partitioning in this experiment. That is, we compared the runtimes for unpartitioned and partitioned tables to show that vertical partitioning is meaningful in most cases. However, as stated above, for pure OLTP workloads an unpartitioned row-store table would be the optimal setup. This aspect of considering table-level decisions and partition-level decisions at the same time will be addressed in the next part of the evaluation.

*Summary.* Our experiments point up that utilizing the store-aware partitioning may further speed up query processing beyond the decision on table level—this will be emphasized in the subsequent experiment. We showed that a meaningful partitioning can reduce workload runtimes although some of the queries are slowed down due to additional joins or unions.

*Combination and Comparison*

In the final experiment, we combined and compared the different approaches for recommending the store. To this end, we used a TPC-H like scenario by using the TPC-H data (with a scale factor of 1) but generating a mixed workload of OLTP queries (inserts and updates for all tables but `nation` and `region`) and OLAP queries (aggregates with and without joins and groupings mainly on `lineitem` and `orders`). We ran this workload for (i) managing all the tables in the row store (*RS only*), (ii) managing all the tables in the column store (*CS only*), (iii) using the recommended store on table level (*Table*), and (iv) using the recommended storage layout including both horizontal and vertical partitioning (*Partitioned*). For a workload containing 5000 queries and a fraction of about 1% of OLAP queries (according to the preceding experiments) the results are depicted in Figure 10. The displayed chart shows that simply managing the whole schema in just one of the two stores results in the highest overall runtimes. With recommendations on table level, the tables `lineitem` and `orders` were put to the column store while the remaining tables have been stored in the row store; this results in clearly lower runtimes compared to *RS only* and *CS only*. For storage recommendations including partitioning the tables `lineitem` and `orders` were partitioned horizontally for newly inserted tuples, and the partition with the historic data of `orders` and the table `part` were partitioned vertically according to update and aggregation attributes. As a result, the overall runtime further decreases by about 40% compared to *Table* or about 65% compared to *CS only*.

*Summary.* With the final experiment, we have shown that it is important to carefully choose the respective store for each table of a given schema—managing the tables in the optimal store results in significantly lower runtimes. By additionally taking partitioning into account the query runtimes can again be significantly decreased.

## 6. RELATED WORK

In this section, we provide an overview of related work in the field of database optimization and hybrid-store data management.

As in the past nearly all of the commercially used databases have been row-oriented, research and effort for speeding up query processing by means of the physical design focused on that very storage paradigm. On the one hand, additional structures like indexes [7] and materialized views [10] as well as their combination [2] were proposed. On the other hand, different kinds of organizing the data have been considered, like horizontal partitioning [3], vertical partitioning [1], and multi-dimensional clustering [17] of the data. In contrast, column-oriented databases are mainly used in data-intensive application fields; here, access patterns typically involve scanning over complete tables. As a consequence, physical design techniques that can be used to speed up row-oriented databases are not appropriate for this kind of storing scheme. Thus, the means for speeding up query processing are mainly compression techniques like the compression of horizontal partitions by tuple frequency proposed by Raman et al. [15] or the compression techniques for BI accelerators described by Lemke et al. [12]. Fortunately, all the techniques mentioned for row stores and column stores for reducing query response times can continued to be used in hybrid-store databases.

There are also approaches to increase query execution efficiency by combining the benefits of both stores. The PAX approach [4], for example, optimizes the typical row-oriented storage by applying a column-oriented layout within memory pages. With the help of this technique they could significantly improve the query execution of range selection, update and OLAP queries. Other solutions like the RCFile approach [11] adopt the idea of the PAX system, but optimize the storage for specific use-cases. In the case of RCFile the data placement is optimized for data warehouses that follow the map reduce scheme. The X100 query engine [6] of the MonetDB column-store database system [5] optimizes the column-oriented storage for high-volume queries by introducing a vectorized data processing. There, they divide the data columns in horizontal chunks (vectors) that better serve the needs of modern CPUs such as loop pipelining. With the help of their approach they showed order of magnitudes faster query execution times on the TPC-H benchmark. While these hybrid approaches greatly increase the query response times, they directly optimize the storage architecture and the data placement on the disk or within main memory. Thus, it is not possible to directly choose the data store with respect to the workload characteristics or the querying behavior. As a result, while these techniques can be used to enhance the query execution efficiency of pure row- or column stores, they are not applicable to actual hybrid stores like the SAP HANA database that offer both stores simultaneously.

An approach that actively decides whether to execute queries on a row- or column store is the Fractured Mirrors approach [14]. It is inspired by the RAID-1 storage technology and maintains two copies of the database—one column-oriented and one row-oriented copy. The query optimizer decides, based on the query characteristics, whether to execute the query on the row- or column-store copy of the data. Thus, rather than optimizing the storage of the data, based on the querying behavior, the database is replicated, doubling the amount of data to be stored. In contrast, our solution deals with a new dimension of the physical database design problem. As hybrid-store databases offer both row and column store, it is required to decide in which store to maintain the data most efficiently for the characteristics of the workload at hand.

The first solution that addresses the optimization of hybrid-store databases is HYRISE [9]. Here, a layout manager partitions tables into vertical partitions based on the expected number of cache misses for a given workload. While HYRISE only considers vertical partitions of tables and is restricted to only a small set of operators, our approach provides recommendations for entire tables as well as for horizontal and vertical partitions of tables and supports all common database operators. The actual goal of HYRISE is to accurately estimate the number of cache misses. As the estimation highly depends on the implementation of the individual operators, the HYRISE approach is very specific to that system. In contrast, our approach directly estimates the runtime of a given workload based on query and data characteristics; the actual cost estimation can be adjusted to the specifics of the individual hybrid-store database system e.g. by benchmarking the system.

## 7. CONCLUSIONS AND FUTURE WORK

In this paper, we introduced the storage advisor, a tool for optimizing the data storage of hybrid-store database management systems to reduce the overall query execution times. While hybrid-store databases combine the advantages of row and column stores, they require database administrators to decide in which store the data should be managed. The storage advisor supports this decision process by analyzing the database schema and a given workload and providing recommendations for storing a table row- or column-oriented. Besides recommendations on the table level, the storage advisor also supports more fine-grained decision by means of considering horizontal and vertical partitioning of a table. The recommendations are provided on the basis of estimated query execution costs that are calculated with the help of our proposed cost model. The cost model is initialized prior to operationally querying the database schema, taking the used hardware and the specifics of the targeted SAP HANA database instance into account. In addition, the storage advisor provides constant support to the database administrator through also offering an online mode. There, the executed workload is recorded and serves together with extended table statistics as input for an continuous evaluation of the currently employed schema configuration. Whenever a more beneficial configuration is found, the necessary adaptations are presented to the database administrator or are directly applied to the database system. Our evaluation shows that with the help of our storage advisor, one can effectively exploit the advantages of both stores and significantly reduce the query execution times on hybrid-store databases. Thus, with the storage advisor we provide important support for administrators of hybrid-store databases, relieving them from the need to manually decide about a fine-grained data storage.

The next steps of the storage advisor work include a fine-grained analysis of the trade-off between the recommendation quality and the amount of statistics to be recorded. Recording very detailed statistics about the workload is very expensive and may neutralize or even outweigh the performance improvements achieved with an optimal storage layout. So, the goal is to identify a preferably small set of statistics that still provides high quality recommendations. Another aspect that we will address is the system load. Here, we will analyze the impact of (heavily) concurrent system accesses and extend the cost model accordingly.

## 8. ACKNOWLEDGMENTS

The work presented in this paper has been carried out in the projects *MIRABEL* and *CUBIST* that are funded by the European Commission under the 7th Framework Programme of ICT.